\documentclass[preprint,nofootinbib]{revtex4-1}
\usepackage{graphicx}
\usepackage[colorlinks]{hyperref}
\usepackage{color}
\usepackage[caption=false]{subfig}
\begin{document}

\title{Investigation of Neutralino Pair Production in Photon-Photon Collider at ILC}
\author{Nasuf SONMEZ}
\affiliation{Ege University, Physics Department, Izmir, Turkey}
\email{nsonmez@cern.ch}
\date{\today}

\begin{abstract}

Neutralino pair production via photon-photon collisions is analyzed in the context of Minimal Supersymmetric Standard Model at future linear collider.
Since photon does not have self coupling, this process is only possible at Next-to-Leading order and all the possible terms are calculated for the photon-photon interaction, including box, triangles and quartic coupling diagrams.
Numerical analysis of the production rates for $ \tilde{\chi}^0_1 \tilde{\chi}^0_1$, $ \tilde{\chi}^0_1 \tilde{\chi}^0_2$  and $ \tilde{\chi}^0_2 \tilde{\chi}^0_2$ are presented for four new distinct benchmark models which are presented in the light of LHC8.
Angular dependence of each neutralino pairs for the benchmark points are also presented.
Total integrated photonic cross section goes up to $1.23\;fb$ and $1.26\;fb$ for the $ \tilde{\chi}^0_1 \tilde{\chi}^0_1$ and $ \tilde{\chi}^0_2 \tilde{\chi}^0_2$ pairs, respectively for the Radiatively driven natural susy benchmark point.
\end{abstract}

\maketitle


\section{Introduction}

After discovering the Higgs particle in LHC\cite{Aad:2012tfa}\cite{Chatrchyan:2012ufa}, one important step is achieved for the Standard Model (SM). 
However any hint for the theories which try to explain how to stabilize the quadratic divergencies arising in self interaction of Higgs field in SM via quantum fluctuations, as well as shortcomings of SM in grand puzzle of the universe would be the next challenge.
Supersymmetry is a theory which introduces an explanation for the strong and electroweak interactions from Planck scale down to the weak scale by introducing supersymmetric partners for all the SM particles. 
Therefore, it reduces the arising quadratic divergences in self interaction of scalar fields to merely logarithmic ones.
Minimal Supersymmetric Standard Model (MSSM)\cite{Martin:1997ns} is a minimal extension of the SM which conceives the supersymmetry.
According to the theory, if R-parity is conserved among super partners, the lightest supersymmetric particle becomes a typical candidate for weakly-interacting dark matter\cite{Martin:1997ns}.
Apparently, the dark matter makes more than one-forth of the energy density of the universe we live in.
The lightest supersymmetric particle in this context becomes neutralino ($\tilde\chi_1^0$) and it escapes detection in the detector. 
It could only be tracked via missing energies in each event. 
There is still ongoing hunt on supersymmetric particles in LHC.

Besides of LHC, there is an ongoing effort for the future International Linear Collider (ILC) by the particle physics community where $e^+e^-$, $e^-e^-$ and $\gamma e$ collisions are considered.
It is also possible to design an $e^+e^-$ linear collider to operate as a $\gamma\gamma$ collider by extracting Compton backscattered photons.
$\gamma\gamma$-collider is considered as a future option in the center of mass energy $\sqrt{s}=250-1000\;\text{GeV}$ with an integrated luminosity of the order of $100\;fb^{-1}$ yearly \cite{Behnke:2013xla} \cite{Behnke:2013lya}. 
The machine is expected to be upgradeable to $\sqrt{s}=1\;TeV$ with total integrated luminosity up to $300\;fb^{-1}$ yearly. 
In the center of mass system of $\gamma\gamma$ collision the energy is peaked around $0.83\sqrt{s}$.

The main task in ILC would be complementing the LHC results, as well as searching clues in Beyond the Standard Model (BSM) such as supersymmetry.
The machine will be capable of studying the properties of new particles and the interactions it makes.
Linear colliders compared to the LHC, have cleaner background and the signals which exhibits new physics could be more easily resolved from the backgrounds.
Besides, since the photon coupling is the same for all quarks and leptons, it is the same for new particles from BSM too.
Pairs of all species, new and exotic ones will be produced in ILC at similar rates.
Consequently, the ILC is an ideal laboratory to study new physics with much more precise measurements.
In this paper, we study the neutralino pair production rates in a photon collider which has been proposed as an option at 
the ILC.

In the light of physics analyses studies on LHC data in pp collisions at $\sqrt{s}=7\;\text{TeV}$ (LHC7) and $\sqrt{s}=8\;\text{TeV}$ (LHC8) with $5\;fb^{-1}$ and $20\;fb^{-1}$ total luminosity respectively, strong limits on SUSY parameter space and constraints on the mass of the sparticles are set \cite{Khachatryan:2014mma}  \cite{Khachatryan:2014qwa} \cite{Khachatryan:2014doa}
.
Using the LHC8 and LHC7 data ATLAS and CMS collaborations, recently, have excluded a very favorable model known as CMSSM for the following cases \cite{CMS:zxa}; i.) $m_{\tilde{g}} \leq1500\;\text{GeV}$ and $m_{\tilde{g}}\simeq m_{\tilde{q}}$, and ii.) $m_{\tilde{g}} \leq1000\;\text{GeV}$ and $m_{\tilde{g}}\ll m_{\tilde{q}}$.
In addition to that, gauge and anomaly mediated SUSY breaking models will also be dismissed if the discovered Higgs boson with mass of $125\;\text{GeV}$ is figured out to be the supersymmetric light CP-even Higgs boson.
All these recent results on susy search put pressure to move on from previously defined popular models to new ones which have interesting signatures
.

In this study, the numerical calculation will be presented for the following benchmark points which are proposed in the light of LHC7 and LHC8 data analysis \cite{Baer:2013ula}; 
i.) Radiatively driven natural SUSY (RNS), 
ii.) mSUGRA/CMSSM, 
iii.) Brummer Buchmuller (BB) benchmark and 
iv.) Natural Susy (NS), respectively. 
For these benchmark points the mass of neutralinos and charginos\footnote{Another study where the chargino pair production for these benchmark point is in preparation.} are at sub-TeV range, which makes them accessible at the ILC in $\gamma\gamma$ collision mode as well.
Even though, all these benchmark points are outside of the limits presented by LHC8, therefore, employing these benchmark models for neutralino pair production could show the potential of the ILC concerning the SUSY searches and possible future optimization for accelerator and detector design.

Neutralino pair production rates in $\gamma\gamma$ collider in the context of supersymmetry, especially, helicity nature of the cross section are
studied before by G.J.Gounaris et all \cite{Gounaris:2003ti}, the distribution of the cross section is comparable to our results.
However, the total integrated photonics cross section 
for the generic Feynman diagrams presented by F.Zhou et all \cite{Zhou:2000bk} are higher than our numerical results at the order of two to three.
In this study we have calculated the neutralino pair production rates including whole set of all possible one-loop level Feynman diagrams, 
the total cross section as a function of center-of-mass (cms) energy for the neutralino pairs via unpolarized photon-collisions are presented.
In addition to these, the angular distribution of the cross section is calculated for the unpolarized photon-collisions 
and evaluation of all these numerical calculations are done for the new benchmark models introduced in \cite{Baer:2013ula}. 

The numerical evaluation of the process $(\gamma\gamma\rightarrow \tilde{\chi}_i^0\tilde{\chi}_j^0)$ have been performed using the 
 packages 
\texttt{FeynArts}\cite{Kublbeck:1992mt,Hahn:2000kx} to generate the Feynman diagrams and corresponding amplitudes, 
\texttt{FormCalc}\cite{Hahn:2006qw} to simplify the fermion chains then square the corresponding amplitudes,
and \texttt{LoopTools}\cite{Hahn:1998yk} to perform the evaluation of scalar and tensor one-loop integrals.  
Due to the complexity of the diagrams it is not illuminating to give the lengthy expressions of the full amplitude.
Instead, we have choose to release the numerical code of the total cross section defined in consequent chapters for the $(\gamma\gamma\rightarrow \tilde{\chi}_i^0\tilde{\chi}_j^0)$ scattering process in terms of susy parameters at the electroweak scale such as $M_1$, $M_2$, $\mu$, $\tan\beta$, mixing angles and sparticle mass spectrum\footnote{URL will be available soon.}.
Thus by providing all these parameters it is possible to calculate the total photonic cross section as well as convoluted one with the photon structure function to give the total cross section in $\gamma\gamma$ collision in $e^+e^-$ colliders.

The content of this paper is organized as follows. 
In Sec-\ref{sec:2}, the neutralino sector in MSSM is discussed. 
In Sec-\ref{sec:3}, analytical expressions regarding the kinematics of the scattering, the total cross section and the convolution of the cross section in $e^+e^-$ machine are given.
In Sec-\ref{sec:4}, numerical results of the total cross section for each benchmark point we chose are discussed.
At last the conclusion is drawn in Sec-\ref{sec:5}.


\section{Expressions for The Neutralino Sector}
\label{sec:2}
According to the MSSM Lagrangian, the mass eigenstates namely the neutralinos are the linear combination of neutral gauginos ($\widetilde{B}$, $\widetilde{W}^3$) and the neutral part of Higgsino fields ($\widetilde{H}_1^0$, $\widetilde{H}_2^0$). 
The relevant part in the Lagrangian responsible for neutralino masses is defined by bilinear fermion field $\psi_i^0=(-i\widetilde{B}, -i\widetilde{W}^3,  \widetilde{H}_1^0, \widetilde{H}_2^0)$ with $i=1,\ldots,4$.
\begin{equation}
\mathcal{L}=-\frac{1}{2}(\psi_i^0)^T\mathcal{M}\psi_j^0+h.c.,
\end{equation}

The neutralinos are denoted by $\tilde{\chi}^0_i~(i=1,\ldots,4)$ and the mixing is determined by the neutralino mixing matrix
\begin{equation}
     \mathcal{M}_{} =  
\left(%
  \begin{array}{lllll}
     M_1				&  0					& -m_Zs_{W}c_{\beta} 	&  m_Z s_{W}s_{\beta}	\\
     0					&  M_2         			&  m_Zc_{W}c_{\beta} 	& -m_Z c_{W}s_{\beta}	\\
    -m_Z s_{W}c_{\beta} 	&  m_Z c_{W}c_{\beta}	& 0					& -\mu 				\\
     m_Z s_{W}s_{\beta} 	& -m_Z c_{W}s_{\beta}	& -\mu 				& 0
  \end{array} 
  \right)
  \label{eq:neumix}
\end{equation}

In R-parity conserved MSUGRA model, the mass matrix depends on some SM parameters such as mass of Z-boson, the weak angle and four unknown parameters which are gaugino mass parameters $M_1$ associated with the $U(1)$ symmetry group, $M_2$ associated with the $SU(2)$, supersymmetric Higgs mass parameters $\mu$ and the last parameter is the ratio of the vacuum expectation values of two Higgs fields $\tan\beta=v_2/v_1$. 
Therefore, $c_W$ and  $s_W$ represents the cos and sin of weak angle respectively, the same is holds true for $c_\beta$ and $s_\beta$. 
The mass parameters in neutralino mass matrix could be complex for CP non-invariant cases, since we ignore the CP violation we took all the parameters as real.


Since the neutralino mass matrix $M$ is Hermitian the eigenvalues are guaranteed to have a real values. 
Thus, it can be diagonalized by an unitary matrix $N$ such that
\begin{equation}
\mathcal{M_D}=N^{*}\mathcal{M}N^{-1}=diag(m_{\tilde{\chi}^0_1},m_{\tilde{\chi}^0_2},m_{\tilde{\chi}^0_3},m_{\tilde{\chi}^0_4}).
\end{equation}

Eigenvalues of the diagonal matrix $\mathcal{M_D}$ are real but not necessarily positive. 
Therefore, it is customary to define mass eigenstate fields (the neutralino masses) with positive values and increasing mass $m_{\tilde{\chi}^0_1}<m_{\tilde{\chi}^0_2}<m_{\tilde{\chi}^0_3}<m_{\tilde{\chi}^0_4}$. 
There are analytical procedures to find the unitary matrix $N$, therefore, usually it is calculated numerically.
Neutralino mass matrix can be diogonalized by real matrix instead of an unitary one, but sometimes mass values could be negative. 
In calculation, those physical mass states which corresponds to negative mass values need to be modified by chiral rotation.
Instead of modifying the negative mass corresponding mass states, we used the Singular Value Decomposition method \cite{Golub:1996}\cite{Leader:2004} to compute the unitary matrix, in a result all the mass eigenvalues are positive.
Hereafter, if we rotate the gauge eigenstates ($\psi_i^0$) to the physical mass eigenstate basis by the unitary matrix $\tilde{\chi}^0_i=N_{ij}\psi_j^0$ then Lagrangian could be written in terms of diagonal physical mass basis.
\begin{equation}
\mathcal{L}_{m}=-\frac{1}{2}\sum_i \mathcal{M_D}\bar{\tilde{\chi}^0_i}\tilde{\chi}^0_i
\end{equation}

\section{The Calculation of The Cross Section}
\label{sec:3}
 
\begin{figure}[htbp]
\centering
		\includegraphics[width=0.20\textwidth]{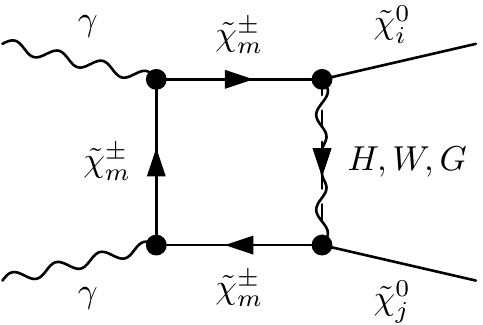}
		\includegraphics[width=0.20\textwidth]{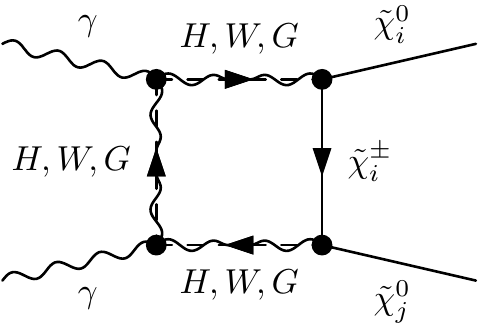}
		\includegraphics[width=0.20\textwidth]{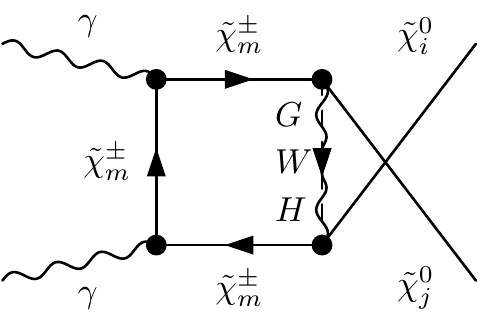}
		\includegraphics[width=0.20\textwidth]{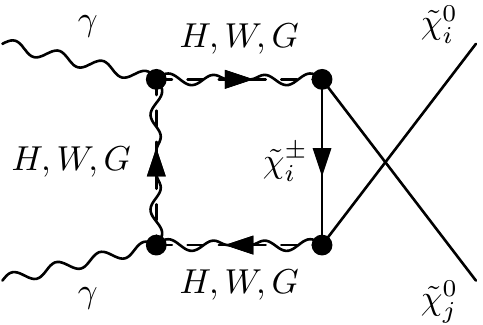}
		
		\includegraphics[width=0.20\textwidth]{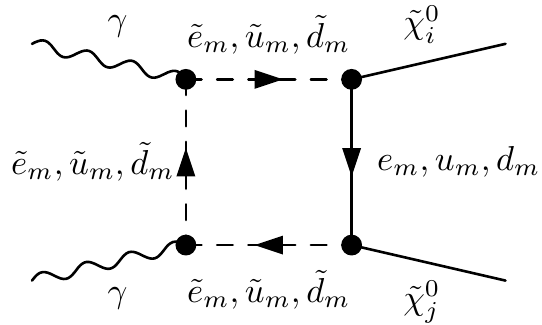}
		\includegraphics[width=0.20\textwidth]{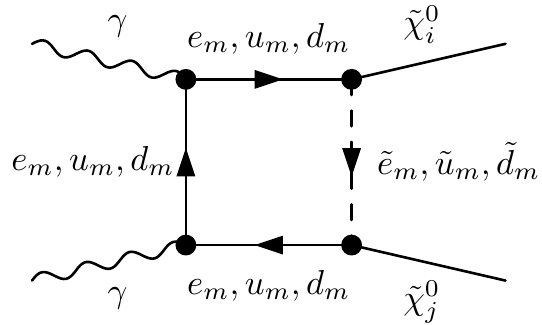}
		\includegraphics[width=0.20\textwidth]{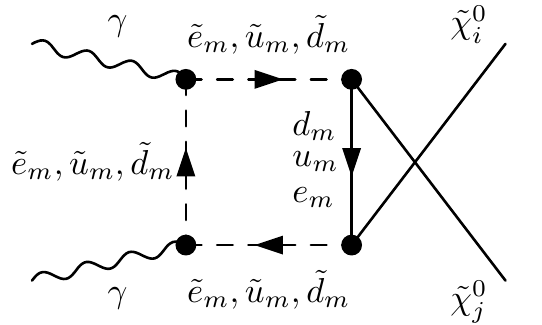}
		\includegraphics[width=0.20\textwidth]{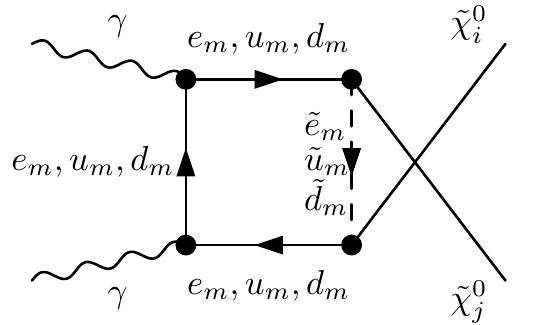}
		\includegraphics[width=0.20\textwidth]{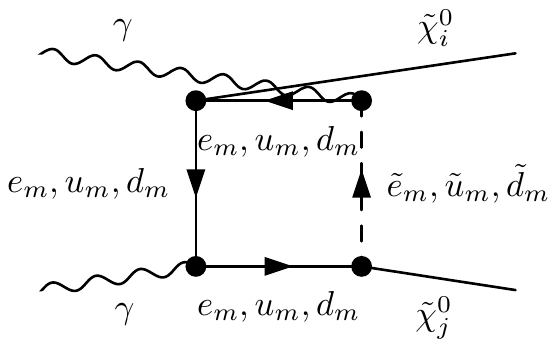}
		\includegraphics[width=0.20\textwidth]{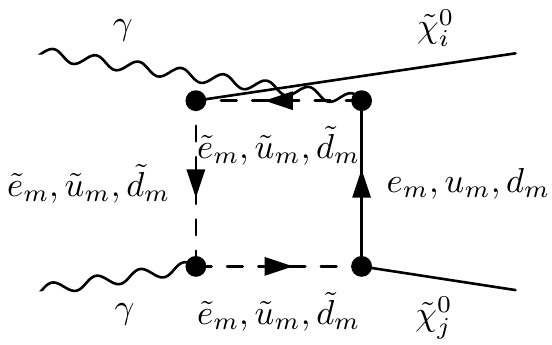}
		\includegraphics[width=0.20\textwidth]{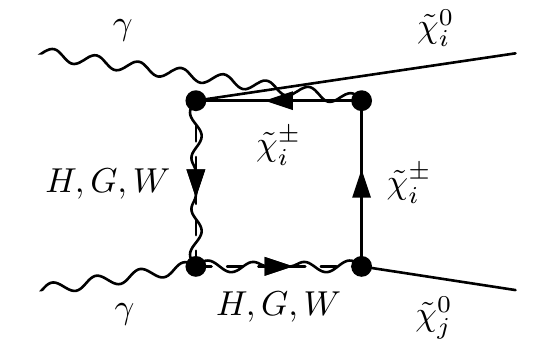}
		\includegraphics[width=0.20\textwidth]{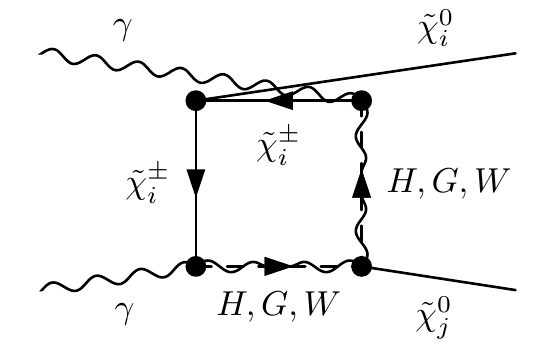}
		\includegraphics[width=0.20\textwidth]{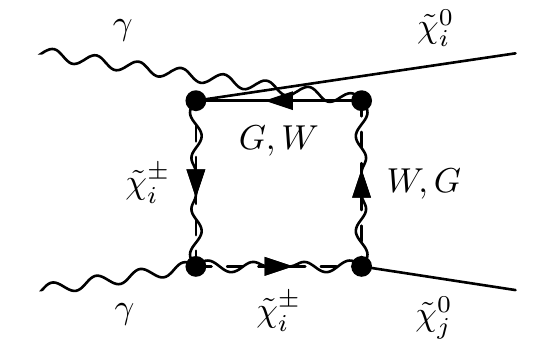}
		\includegraphics[width=0.20\textwidth]{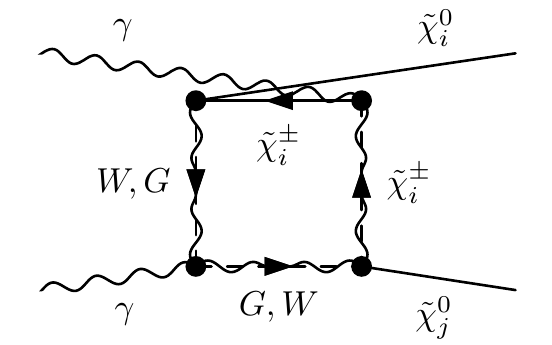}
\caption{One-loop Feynman box diagrams for the neutralino pair production at the ILC via photon-photon fusion.}
\label{fig:neupair_box}
\end{figure}

\begin{figure}[htbp]
\centering	
	\includegraphics[width=0.20\textwidth]{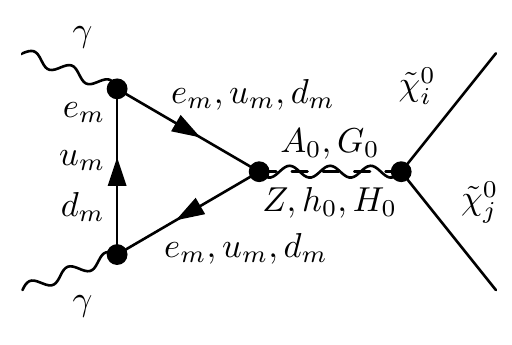}
	\includegraphics[width=0.20\textwidth]{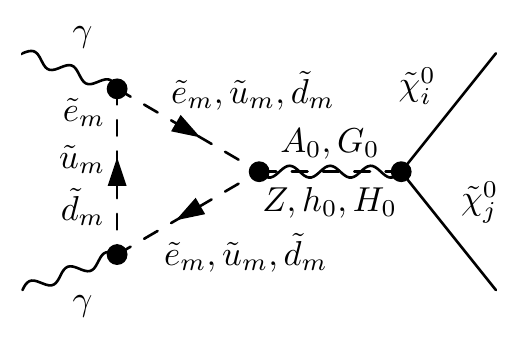}
	\includegraphics[width=0.20\textwidth]{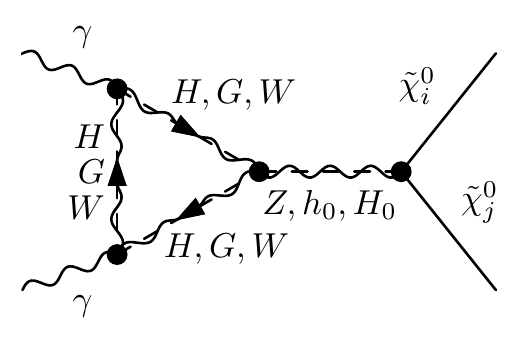}
	\includegraphics[width=0.20\textwidth]{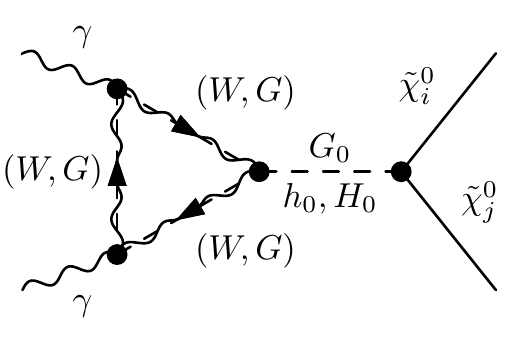}
	\includegraphics[width=0.20\textwidth]{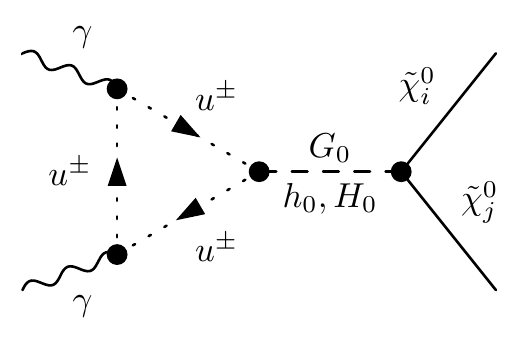}
	\includegraphics[width=0.20\textwidth]{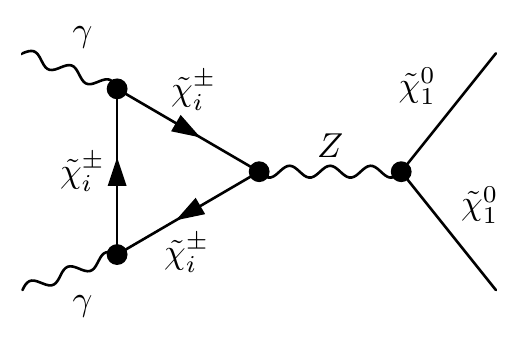}
\caption{One-loop Feynman triangle diagrams for the neutralino pair production at the ILC via photon-photon fusion.}
\label{fig:neupair_tri}
\end{figure}

\begin{figure}[htbp]
\centering
	\includegraphics[width=0.35\textwidth]{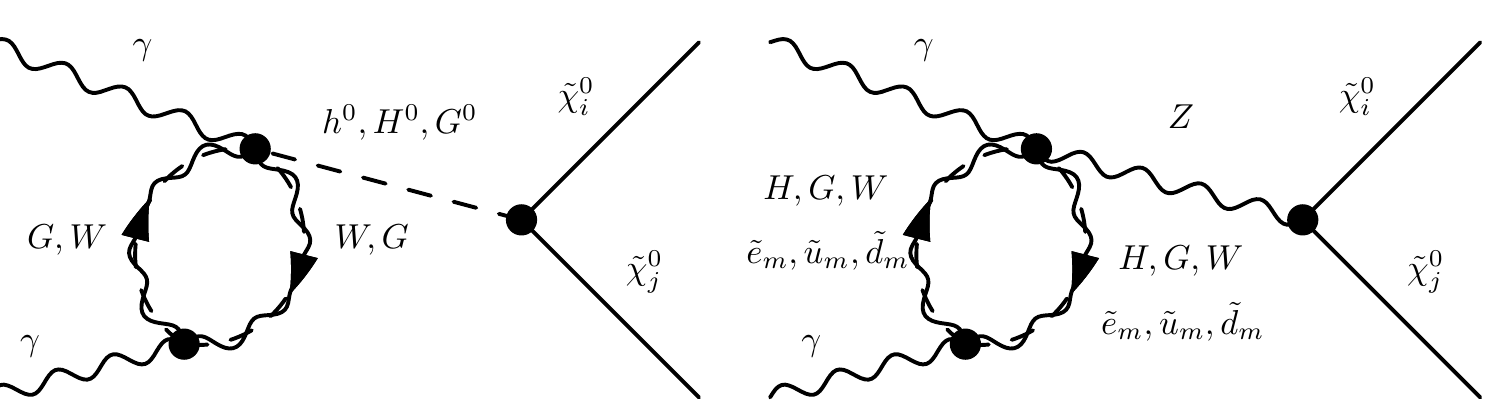}
	\includegraphics[width=0.35\textwidth]{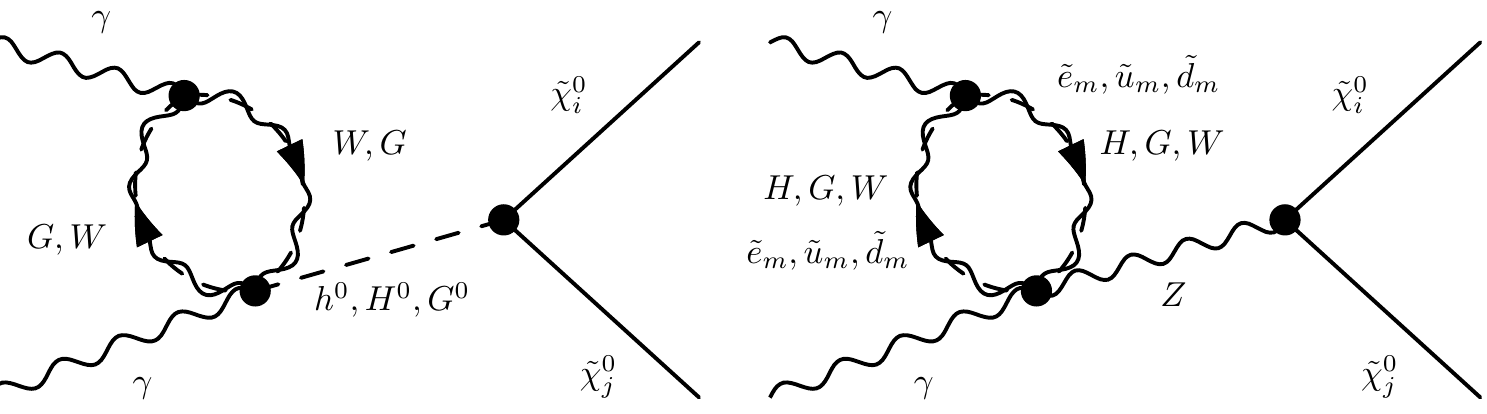}
	\includegraphics[width=0.35\textwidth]{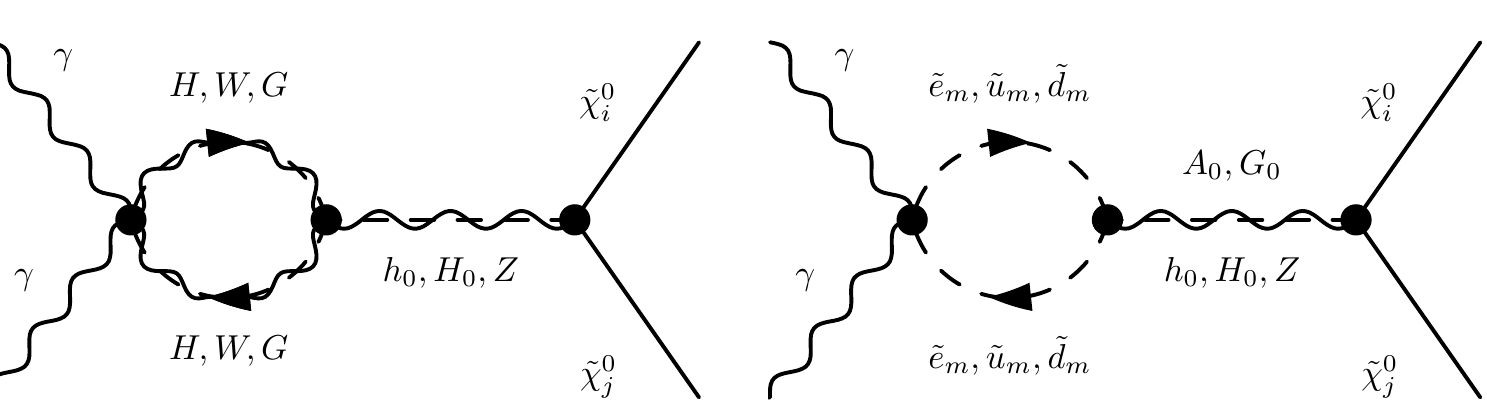}
	\includegraphics[width=0.35\textwidth]{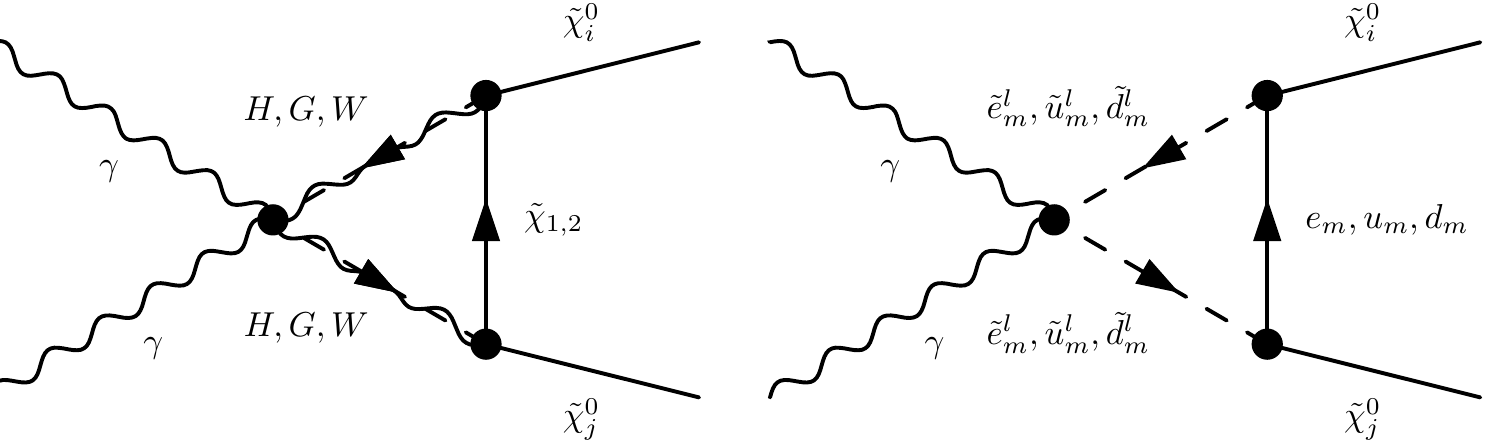}
\caption{One-loop Feynman quartic diagrams for the neutralino pair production at the ILC via photon-photon fusion.}
\label{fig:neupair_quart}
\end{figure}

In this section, analytical expressions of photonic cross section and convoluted cross section with the photon luminosity in $e^+e^-$ collider for neutralino pair production are given.
Throughout this paper, the process for the neutralino pair production at NLO via photon-photon collision is denoted as 
\begin{eqnarray*}
\gamma (k_1,\mu)\;\gamma(k_2,\nu)\;\rightarrow \;  \tilde{\chi}^0_i (k_3)\;\tilde{\chi}^0_j(k_4) \;\;\; (i,j=1,2)\,,\nonumber
\end{eqnarray*}
where $k_a$ $(a=1,...,4)$ are the four momenta of the incoming photons and outgoing neutralinos respectively, whereas $\mu$ and $\nu$ represents the polarization vectors of incoming photons. 
Since photon doesn't have coupling to itself, neutralino pair production via photon-photon collision is only possible at the lowest one-loop level.
All the relevant Feynman diagrams contributing to the subprocess $\gamma \gamma \rightarrow  \tilde{\chi}^0_i \tilde{\chi}^0_j$ at the one-loop level are depicted in Fig-\ref{fig:neupair_box}-\ref{fig:neupair_quart}, they are generated using \texttt{FeynArts}.

The amplitudes are constructed using \texttt{FeynArts}, the relevant part of the Lagrangian and corresponding Feynman rules for the vertices are defined in \cite{Haber:1984rc} and the \texttt{FeynArts} implementation itself is given in \cite{Hahn:2001rv}.
 

In terms of loop type we could classify the one-loop diagrams into three distinct groups, named as box, triangle and quartic diagrams.
In Fig-\ref{fig:neupair_box}, all the possible box diagrams for the process is drawn, where the straight lines represents fermonic particles, dashed lines represents the scalar particles, wave lines represents bosonic fields and dashed-wave are either scalar or bosonic nature.
There is another set of box-diagrams not drawn in Fig-\ref{fig:neupair_box} where particles are running in oposite direction in each loop.
The most of the computing time is spent on these box diagrams.

Triangle loop  diagrams for the process are drawn in Fig-\ref{fig:neupair_tri}. 
The SM and squark particles are running in loops in each direction and $h_0,H_0,A_0,G_0$ and $Z$ particles are intermediated between loop and neutralino pairs. 
$(W,G)$ in loop represents the combinatorics of either two W and one G or two G and one W particle is running.
The reverse loop flow is not drawn for triangle diagrams.
The diagrams which have quartic couplings are depicted in Fig-\ref{fig:neupair_quart}.
The particles in loops of the quartic coupling diagrams are running in each direction, as usual.

In calculation we take in to account all the possible one loop level diagrams, therefore, it is not necessary to take in to account the renormalization because ultraviolet divergence is cancelled automatically. 
In this study the calculation is performed in the \emph{'t Hooft-Feynman} gauge where the gauge boson propagators are in simple form so the computations are simplest.

The corresponding Lorentz invariant matrix element for the one-loop level process is written as a sum over box-dagrams (Fig-\ref{fig:neupair_box}), triangle-diagrams (Fig-\ref{fig:neupair_tri}) and quartic ones depicted in Fig-\ref{fig:neupair_quart},
\begin{equation}
{\cal M}= {\cal M}_{box}+ {\cal M}_{tri}+ {\cal M}_{quart}\,,
\end{equation}
where due to the Fermi statistics in the calculation of the amplitudes there is relative $(-1)$ sign between the diagrams obtained by exchanging the neutralinos at the final state in Fig-\ref{fig:neupair_box}, \ref{fig:neupair_tri} and \ref{fig:neupair_quart}.

In numerical calculation, the scattering amplitude is evaluated in the center of mass frame, 
denoting the four-momentum and scattering angle by ($k, \theta$), the energy $(k_i^0)$ and momentum $(\vec{k}_i)$ of the incoming and outgoing particles are given below in terms of cms energy $\hat{s}$ of incoming photons and neutralino masses $m_i^2$:
\begin{eqnarray} \label{eq:cm}
& k_1=\frac{\sqrt{\hat s}}{2}(1,0,0,1),k_2=\frac{\sqrt{\hat s}}{2}(1,0,0,-1),\\
& k_3=(k_3^0,|\vec{k}| \sin\theta,0,|\vec{k}| \cos\theta),\\
& k_4=(k_4^0,-|\vec{k}| \sin\theta,0,-|\vec{k}|\cos\theta)\\
& k_3^0=\frac{\hat s+m_i^2-m_j^2}{2 \sqrt{\hat s}},~k_4^0=\frac{\hat s+m_j^2-m_i^2}{2 \sqrt{\hat s}},\\
& |\vec{k}|=\frac{1}{2 \sqrt{\hat s}}\sqrt{(\hat s-m_i^2-m_j^2)^2-4m_i^2 m_j^2}.
\end{eqnarray}

Sum over the polarization vectors of the incoming photons are calculated using the following vectors:
\begin{equation} 
\epsilon_1^\pm=\frac{1}{\sqrt{2}}(0,\mp1,-i,0),~\epsilon_2^\pm=\frac{1}{\sqrt{2}}(0,\pm1,-i,0).
\end{equation}

After summing over the helicities of the neutralinos and the polarization vectors of the incoming photons, the cross section of the unpolarized photon collisions is calculated by 
\begin{equation}
\hat{\sigma}_{\gamma\gamma\rightarrow \tilde\chi_i^0\tilde\chi_j^0}(\hat{s})=\frac{ \lambda( \hat{s},m_{\tilde\chi_i^0}^2, m_{\tilde\chi_j^0}^2 )}{16 \pi \hat{s}^2} \left(\frac{1}{2}\right)^{\delta_{ij}}\frac{1}{4} \sum_{hel}{|\mathcal{M}|^2}\,,
\label{eq:partcross}
\end{equation}
where 
\begin{equation}
\lambda( \hat{s},m_{\tilde\chi_i^0}^2, m_{\tilde\chi_j^0}^2 )=\sqrt{ (\hat{s}-m^2_{\tilde\chi_i^0}-m^2_{\tilde\chi_j^0})^2-4m^2_{\tilde\chi_i^0}m^2_{\tilde\chi_j^0} }/2
\end{equation}
is simply the K\"allen function for the phase space of outgoing neutralino pairs, the factors $\left(\frac{1}{2}\right)^{\delta_{ij}}$ and $\frac{1}{4}$ are respectively due to the identical-particle at the final state and helicity average of the neutralinos, lastly,  $i$ and $j$ runs over the neutralino flavors at the final state.

The pair production of neutralinos via photon-photon collision is possible by using the laser back-scattering technique on electron beam in $e^+e^-$ at ILC.
The big fraction of the cms energy of the electron beam could be transferred to the photon collision.
Then 
$\gamma\gamma\rightarrow \tilde\chi_i^0\tilde\chi_j^0$ process could be taken as a subprocess in $e^+e^-$ collisions. 
Thus, the total cross section of $e^+e^-\rightarrow \tilde\chi_i^0\tilde\chi_j^0$ could easily be calculated via convoluting the photonic cross section $\hat{\sigma}_{\gamma\gamma\rightarrow \tilde\chi_i^0\tilde\chi_j^0}(\hat{s})$ with the photon luminosity in $e^+e^-$ collider.

The total integrated photonics cross section cross section is defined as 
\begin{equation}
\label{eq:total_cross}
\sigma(s)=\int_{x_{min}}^{x_{max}} \hat{\sigma}_{\gamma\gamma\rightarrow  \tilde\chi_i^0\tilde\chi_j^0}( \hat{s};\; \hat{s}=z^2s ) \frac{dL_{\gamma\gamma}}{dz}\;dz\,,
\label{eq:foldcross}
\end{equation}
where $s$ and $\hat{s}$ are the cms energy in $e^+e^-$ collider and $\gamma\gamma$ subprocess, respectively. $x_{min}$ is the threshold energy for neutralino pair to produce and given as $x_{min}=(m_{\tilde\chi_i^0}+m_{\tilde\chi_i^0})/\sqrt{s}$ and the maximum fraction of the photon energy is taken as $x_{max}=0.83$ \cite{Telnov:1989sd}. 
Therefore, the distribution function of the photon luminosity is defined as
\begin{equation}
\frac{dL_{\gamma\gamma}}{dz}=2z\int_{z^2/x_{max}}^{x_{max}}\frac{dx}{x}F_{\gamma/e}(x)F_{\gamma/e}\left(\frac{z^2}{x}\right)\,,
\end{equation}
where $F_{\gamma/e}(x)$ is the energy spectrum of the Compton back scattered photons from initial unpolarized electrons and it is defined as a function of fraction $x$ of the longitudinal momentum of the electron beam \cite{Telnov:1989sd}.

 
\section{Numerical Results and Discussion}
\label{sec:4}
In this section, some numerical results for the neutralino pair production at one-loop level via photon-photon collisions are presented. 
For the numerical analysis the following input parameters are taken from \cite{Eidelman:2004wy} such as  $m_W = 80.399 \;GeV$ $m_Z=91.1887\;GeV$, $m_t=173.34\;GeV$, $s_W^2=0.2315$ and $\alpha(m_Z)=1/127.934$. 

Apart from SM parameters, we also need to choose a parameter region for the MSSM.
In this study, neutralino pair production is calculated for the benchmark models specifically introduced for the ILC by the constraints set from LHC7 and LHC8. 

A very important feature of supersymmetry is that all three gauge couplings defined in SM meet at one point - this unification is also predicted by GUTs and string theories - which is one of the reason why supersymmetry attracted so much attention.
Besides, SUSY at the weak scale also gives a solution to the so called hierarchy problem.
To reconcile supersymmetry with the experimental results the supersymmetry is broken slightly so that the sparticles have heavier mass spectrum.
The breaking scale, therefore, is closely related to the size of the quantum corrections in scalar sector and it can not be so high at the order of ten TeV.
If supersymmetry breaking scale is at the order of TeV, the sparticles masses will be around weak scale.
However, the results coming from LHC8 make all that nice picture to fade away.
The exclusion of gap between sparticle mass scale and weak scale increases the breaking scale
in a result the so called little hierarch problem \cite{Martin:1997ns} resurrects.
The benchmark points which are mentioned before are introduced to fit into this picture drawn by LHC8.
These benchmark points specifically chosen for having low contribution $(\Delta_{EW})$ to electroweak observable such as Z-mass.

According to \cite{Baer:2013ula}, to achieve low $\Delta_{EW}$, susy-breaking contribution to the Higgs potential $|m^2_{H_u}|$, Higgs-doublet mixing parameter $\mu^2$ and radiative contribution $|\Sigma_u^u|$ all needs to be around $m^2_Z/2$ to within a factor of a few \cite{Baer:2012up} \cite{Baer:2012cf}. 
This implies the following ranges for these parameters:
\begin{itemize}
\item  $\mu$ is favored to be in the $100<\mu<300\;\text{GeV}$ range.
\item $|m_{H_u}|_{\text{weak}}\approx  100-300\;\text{GeV}$. 
\item To minimize the radiative corrections coming from stop $\Sigma_u^u(\tilde{t}_{i})$, it is required to have large stop mixing $A_0\pm \approx1.6m_0$, which also raises the lightest Higgs mass up to the $\sim125\;\text{GeV}$ level.
\end{itemize}
In the light of these points, we carried out the numerical computation and calculated the energy dependence and angular distribution of $\gamma\gamma\rightarrow\tilde{\chi}^0_i\tilde{\chi}^0_j$ cross section for the following benchmark points.
The total photonic cross section for each neutralino pairs are also presented in Table-\ref{tab:photcross} for two distinct cms energy ($\sqrt{s}=0.5\,\text{TeV}-1.0\,\text{TeV}$) in  each benchmark point.\footnotetext{SLHA files for these benchmark points are located at \cite{bib:webpage}.}
More information for each benchmark point could be found at \cite{Baer:2013ula} and references therein.

\begin{figure}[htbp]
\centering
	\subfloat[]{\includegraphics[width=0.48\textwidth]{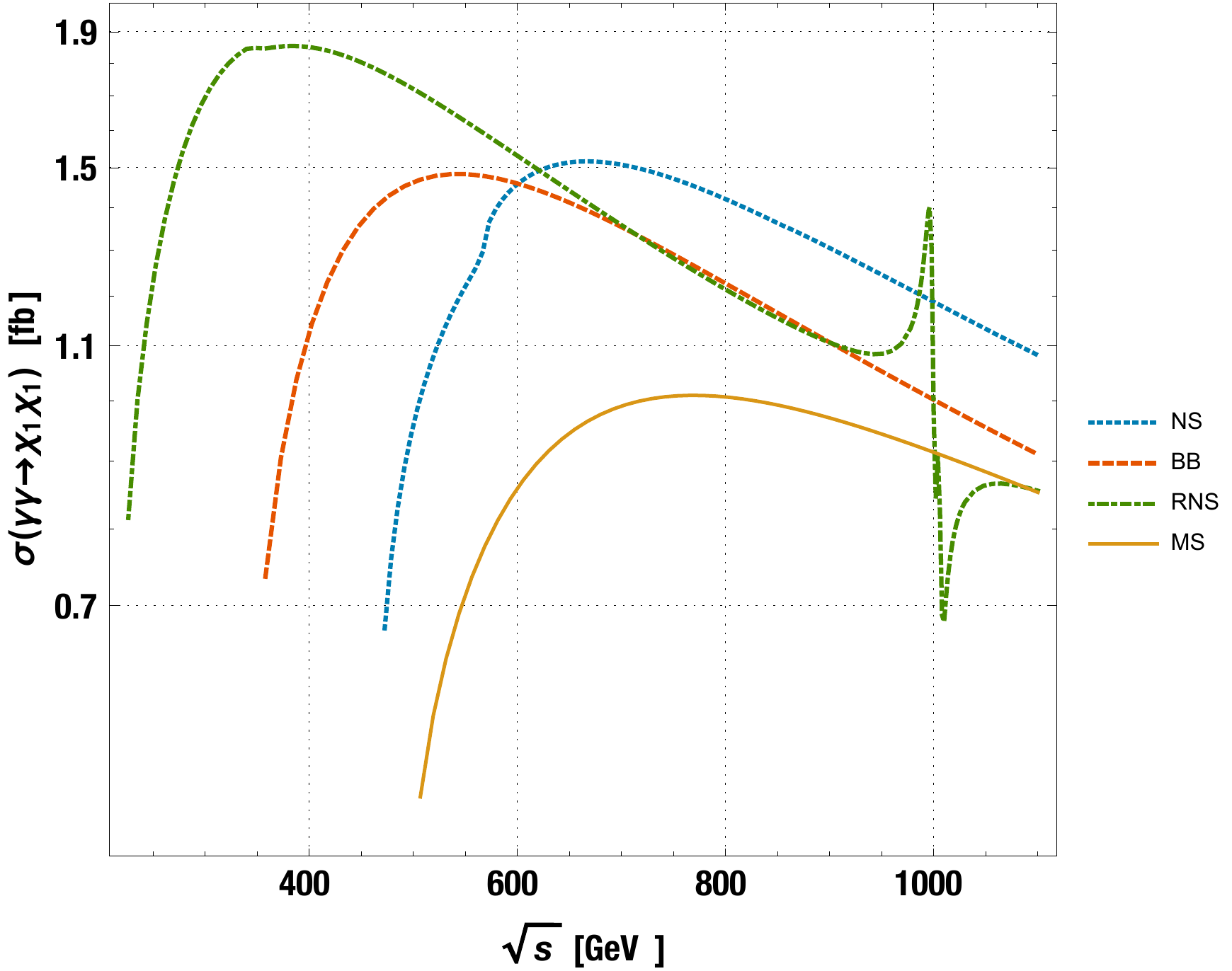}	\label{fig:fig1a}}
	\subfloat[]{\includegraphics[width=0.48\textwidth]{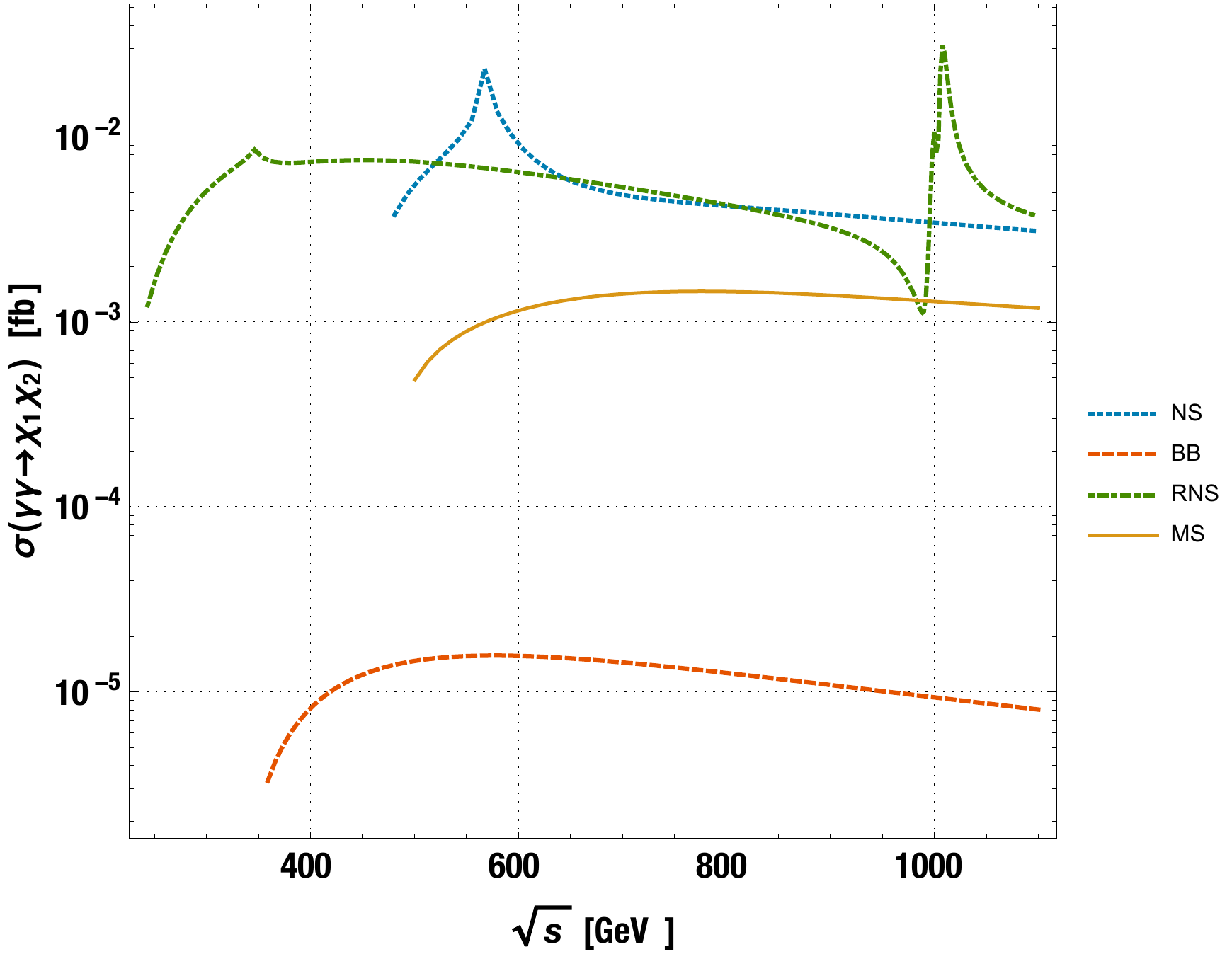}	\label{fig:fig1b}}

	\subfloat[]{\includegraphics[width=0.48\textwidth]{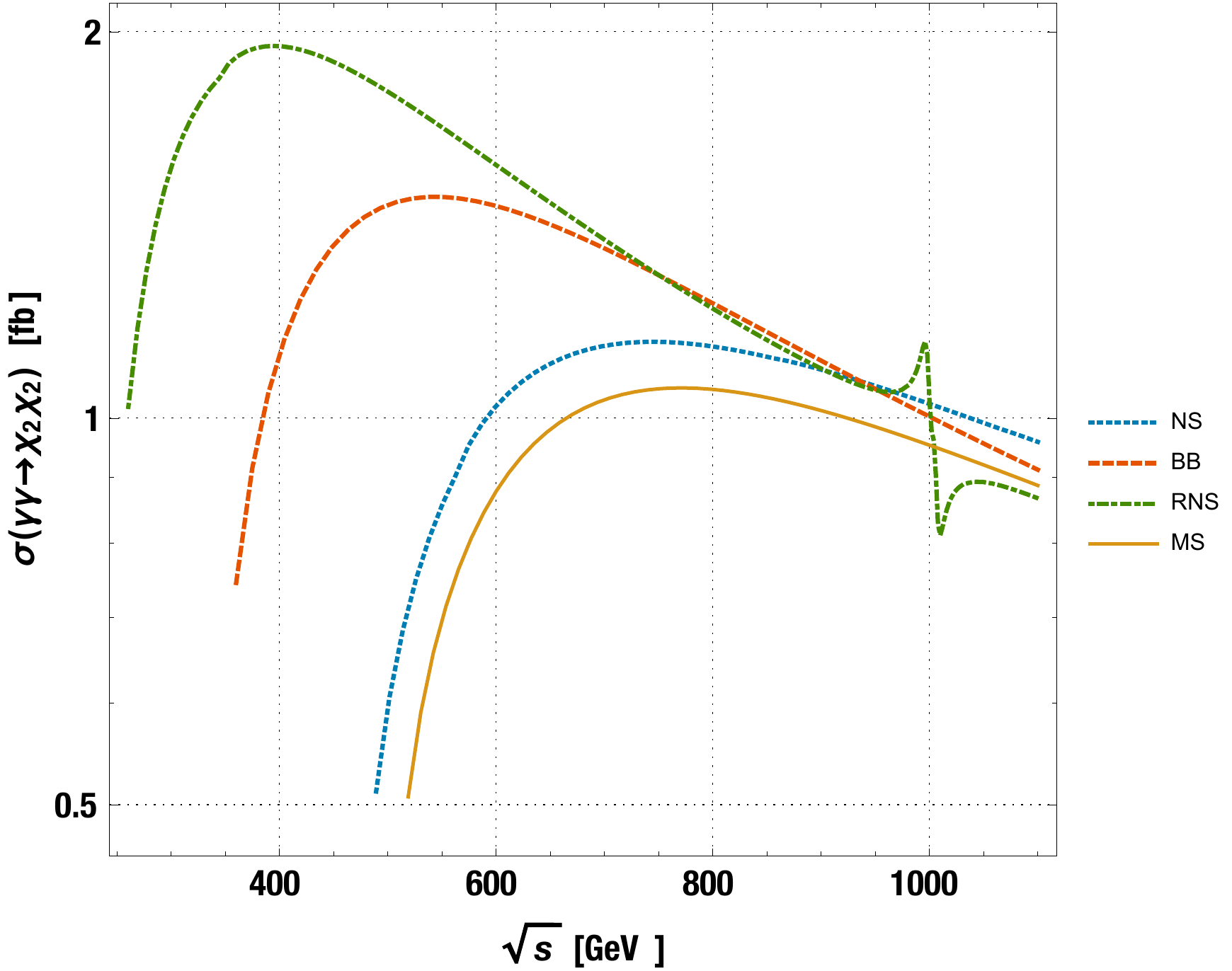}	\label{fig:fig1c}}
\caption{Integrated cross section of the process $\gamma\gamma\rightarrow\tilde\chi_i^0\tilde\chi_j^0$ as a function of $\hat{s}$. a.) $ij=11$, b.) $ij=12$ and c.) $ij=22$. Dotted line stands for NS, dot-dashed line stands for RNS, dashed is for BB and straight line represents mSUGRA benchmark scenarios.}
\label{fig:fig1}
\end{figure}

\begin{figure}[htbp]
\centering
	\subfloat[]{\includegraphics[width=0.48\textwidth]{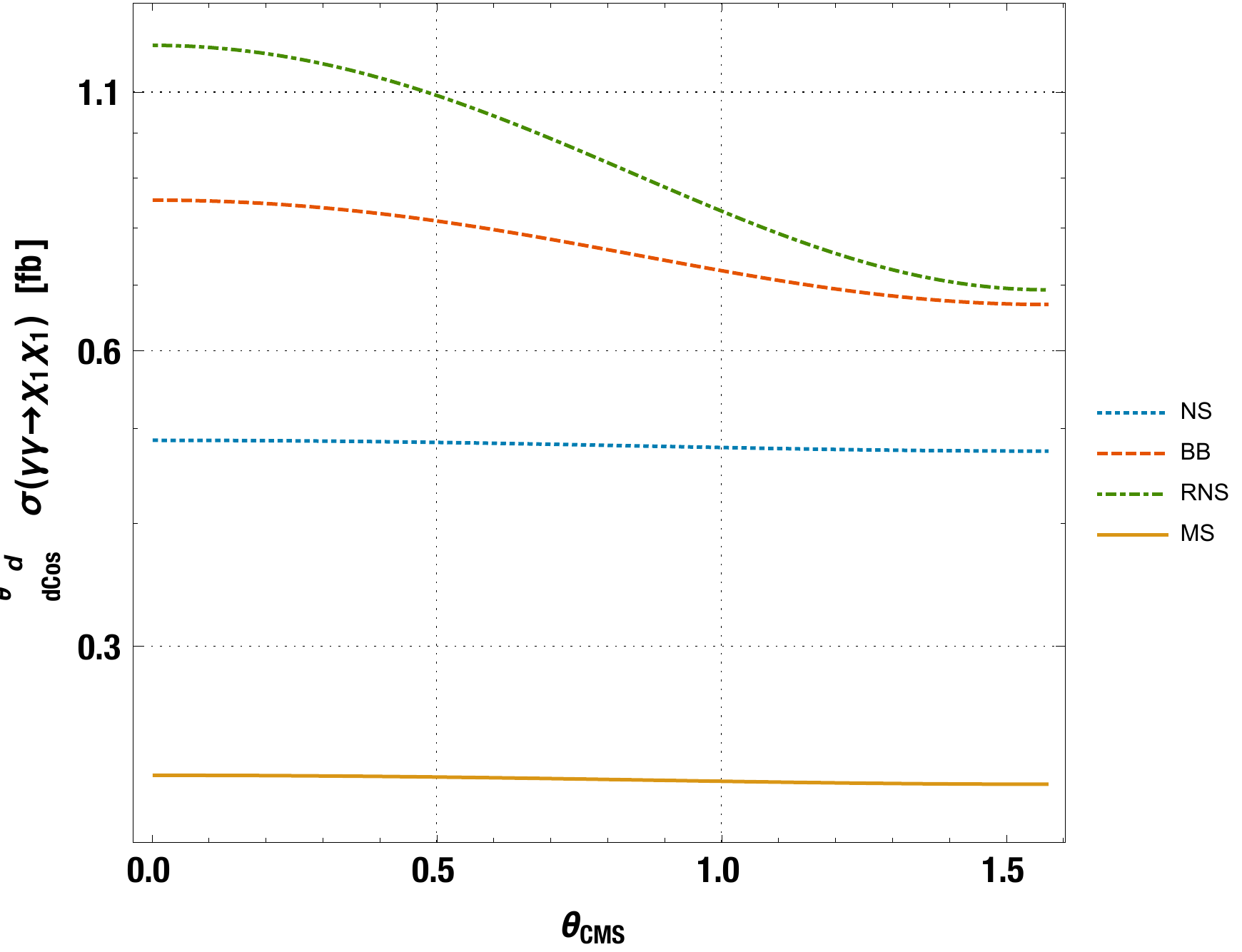}	\label{fig:fig2a}}
	\subfloat[]{\includegraphics[width=0.48\textwidth]{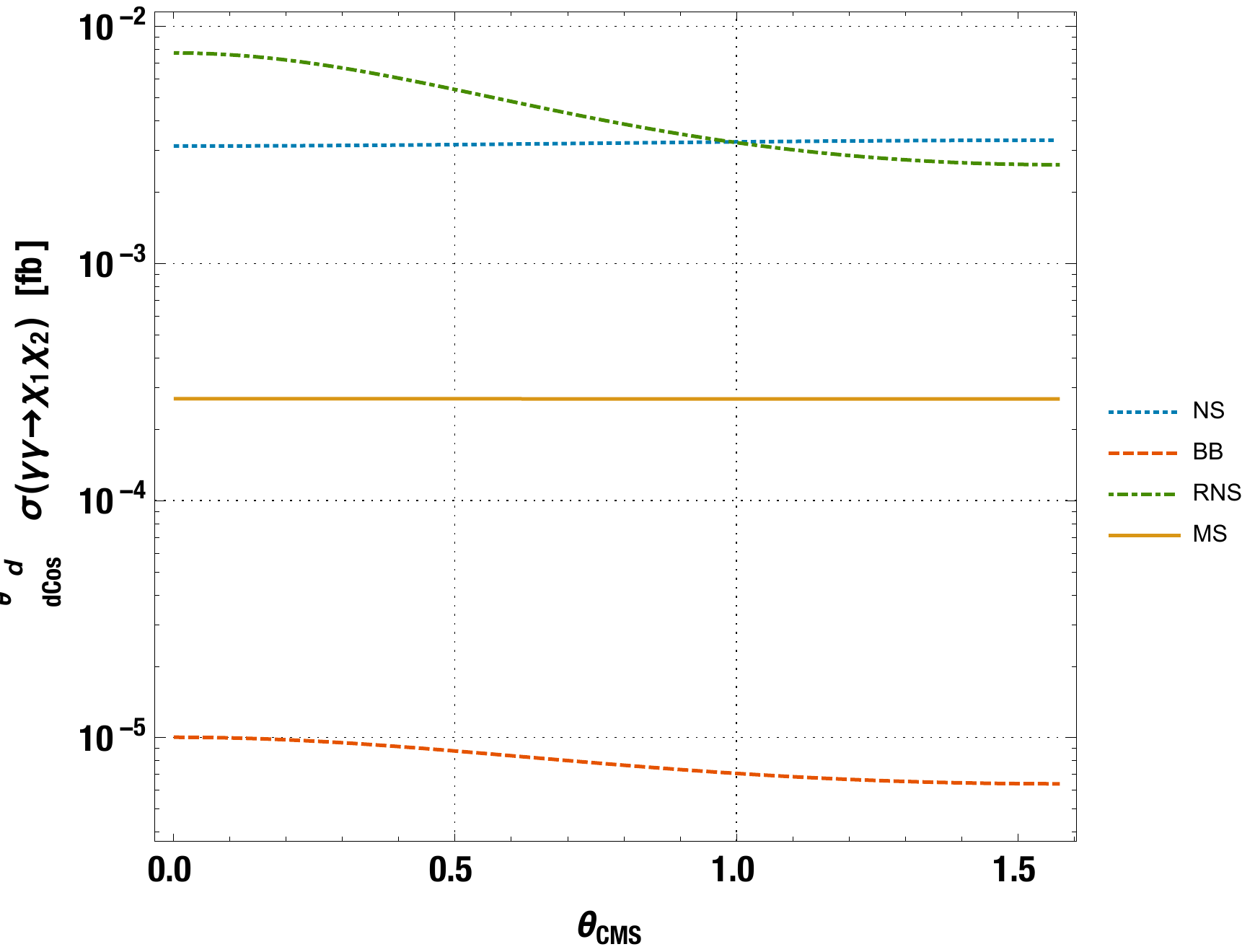}	\label{fig:fig2b}}

	\subfloat[]{\includegraphics[width=0.48\textwidth]{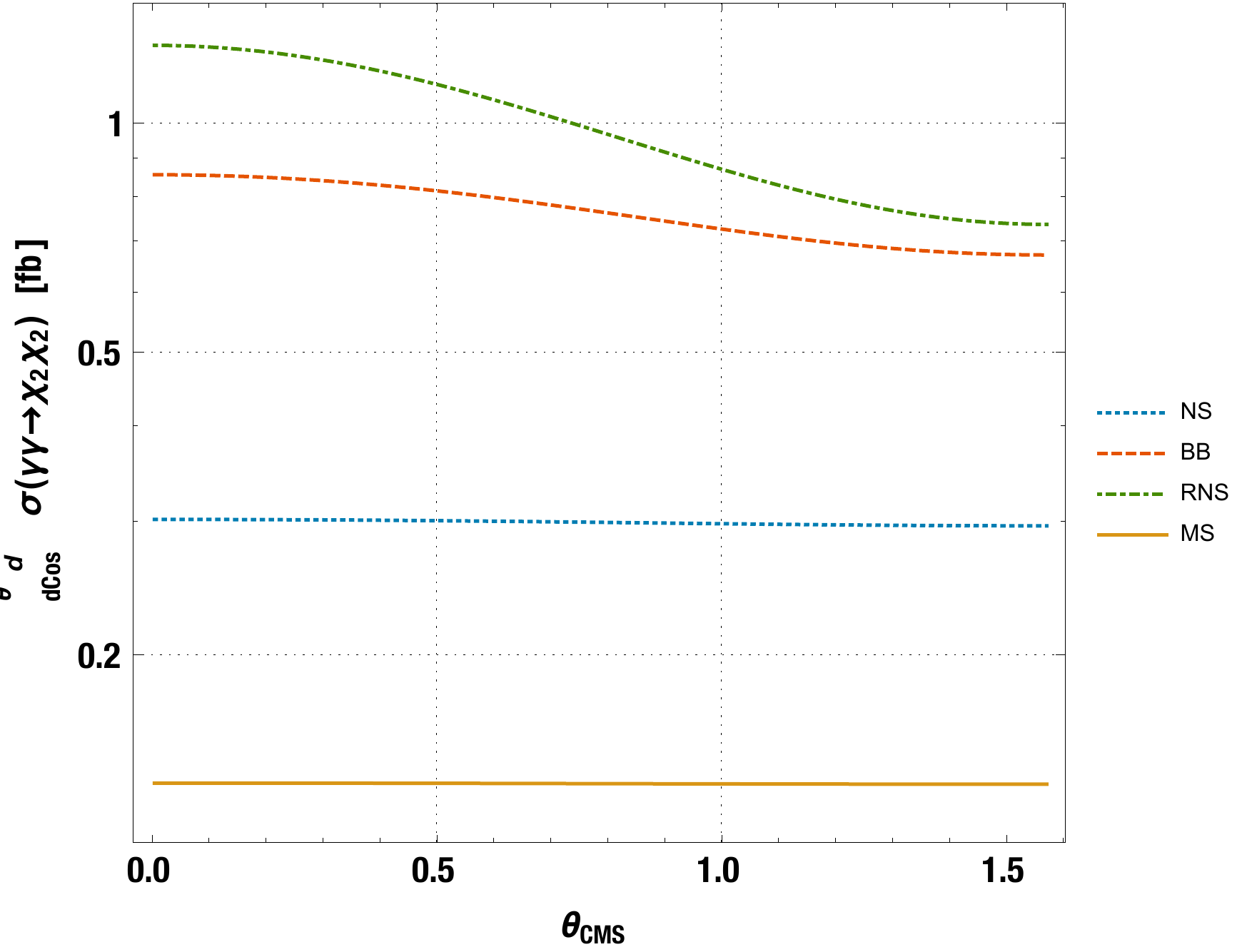}	\label{fig:fig2c}}
\caption{The angular distribution of $\gamma\gamma\rightarrow\tilde\chi_i^0\tilde\chi_j^0$ as a function of $\theta$ in cms for $\sqrt{s}=0.5\;\text{TeV}$. a.) $ij=11$, b.) $ij=12$ and c.) $ij=22$. Dotted line stands for NS, dot-dashed line stands for RNS, dashed is for BB and straight line represents mSUGRA benchmark scenarios.}
\label{fig:fig2}
\end{figure}

\begin{figure}[htbp]
\centering
	\subfloat[]{\includegraphics[width=0.48\textwidth]{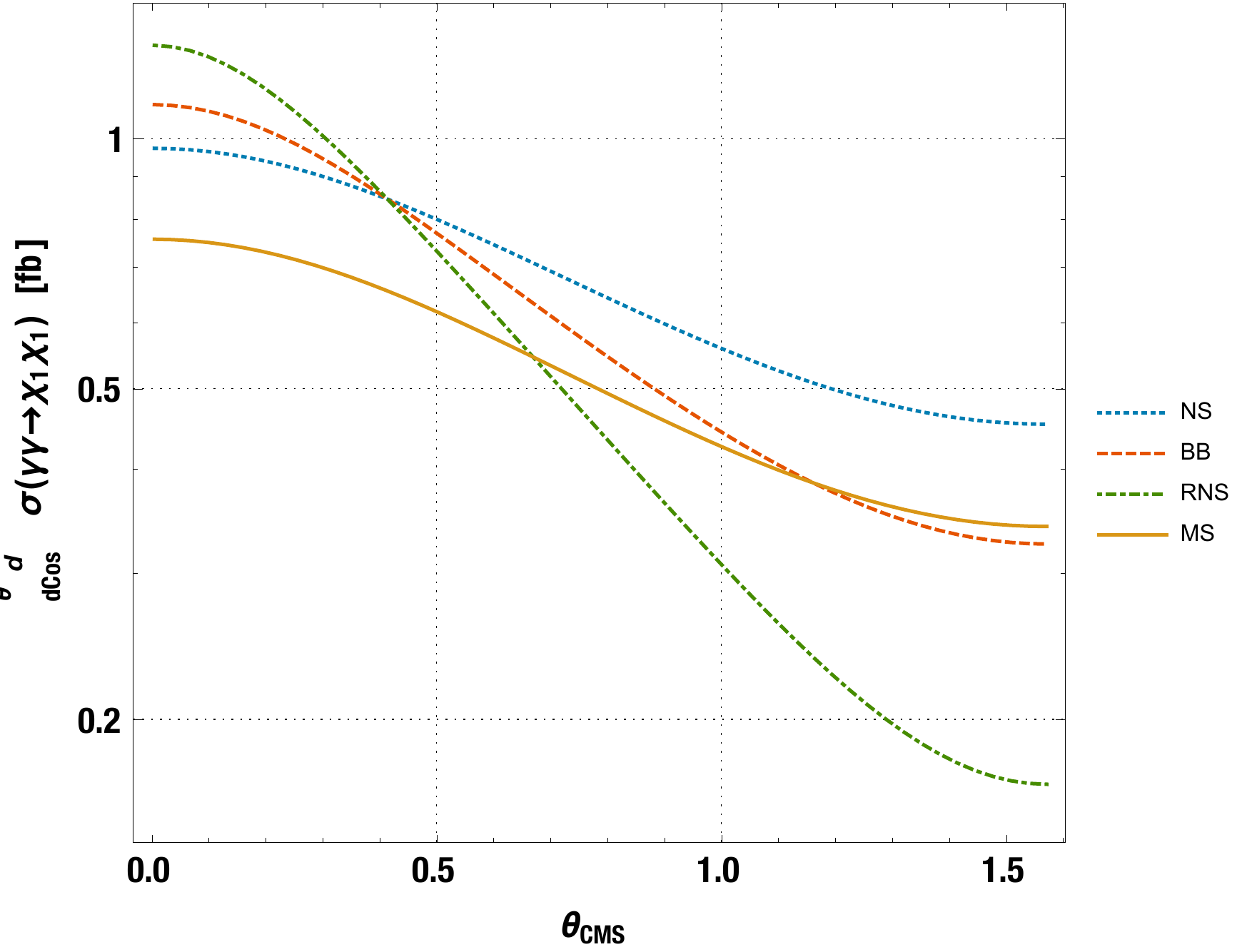}	\label{fig:fig3a}}
	\subfloat[]{\includegraphics[width=0.48\textwidth]{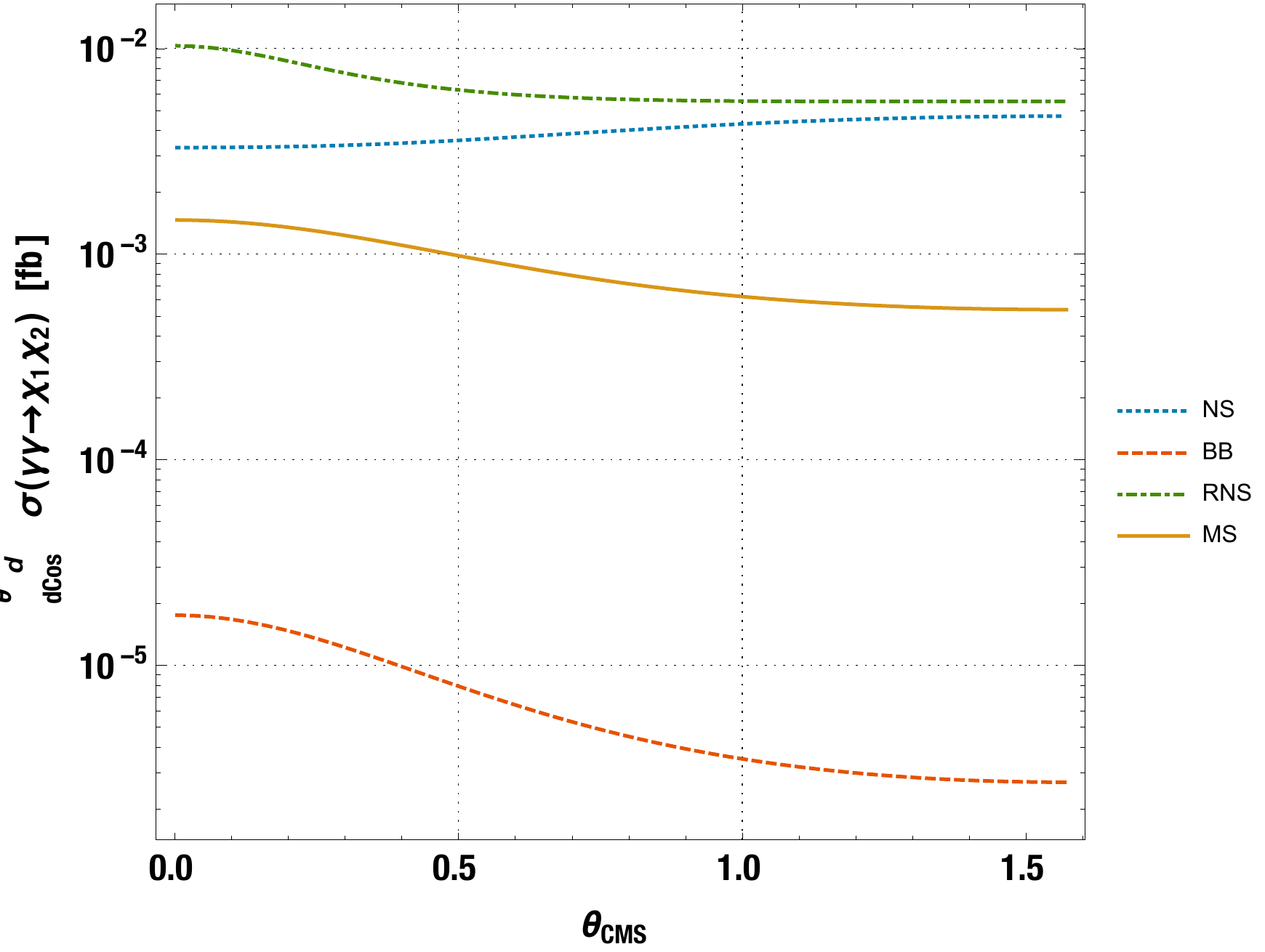}	\label{fig:fig3b}}

	\subfloat[]{\includegraphics[width=0.48\textwidth]{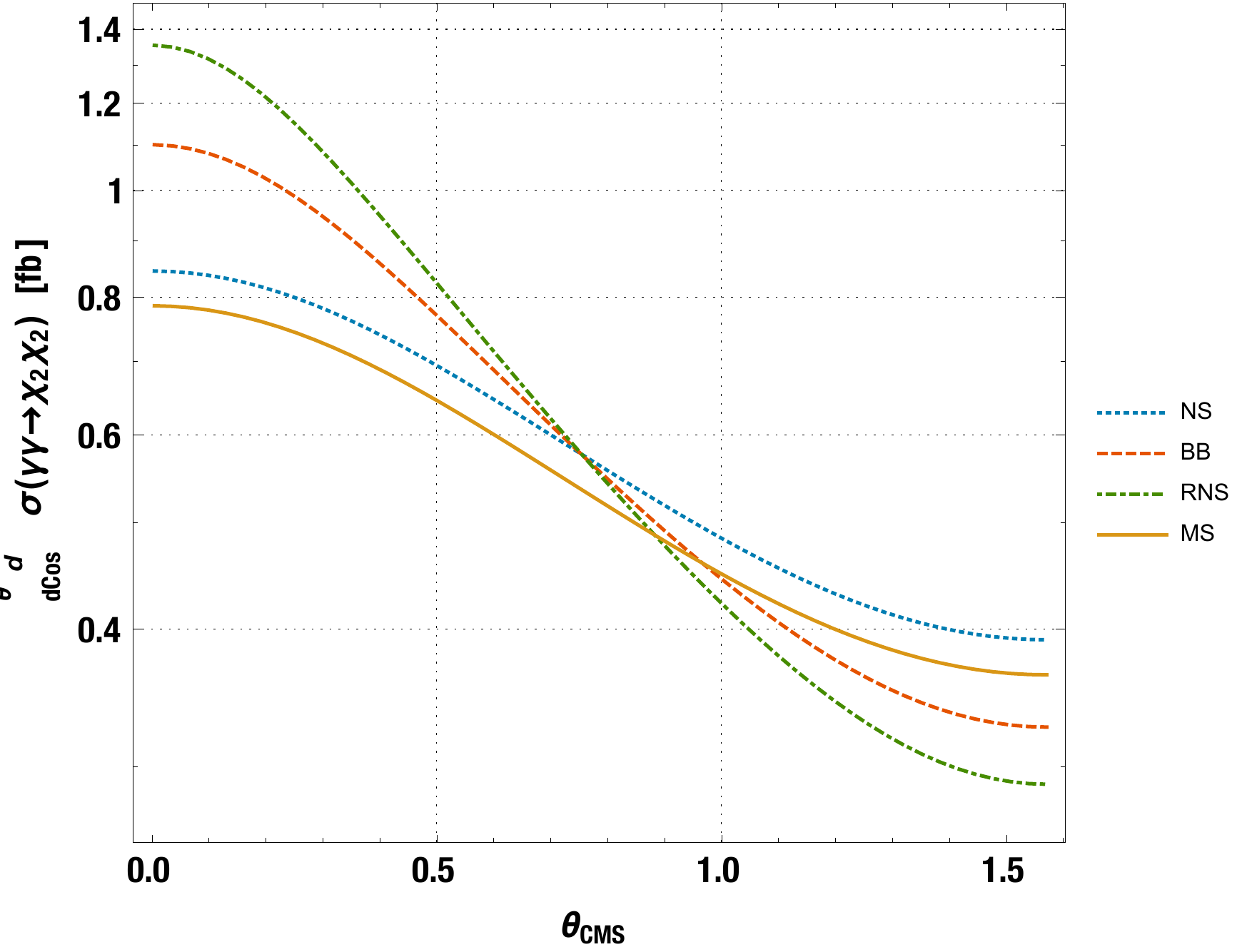}	\label{fig:fig3c}}
\caption{The angular distribution of $\gamma\gamma\rightarrow\tilde\chi_i^0\tilde\chi_j^0$ as a function of $\theta$ in cms for $\sqrt{s}=1.0\;\text{TeV}$. a.) $ij=11$, b.) $ij=12$ and c.) $ij=22$.  Dotted line stands for NS, dot-dashed line stands for RNS, dashed is for BB and straight line represents mSUGRA benchmark scenarios.}
\label{fig:fig3}
\end{figure}

\begin{itemize}

\item \emph{Radiatively driven natural SUSY (RNS)} : 
This model is motivated by minimizing $\Delta_{EW}$, therefore, it still sustains the unification of the gauge couplings and radiative electroweak symmetry breaking.
Minimization of $\Delta_{EW}$ is achieved by requiring Higgs-doublet mixing parameter $\mu$ to be around $\mu\sim100-300\,GeV$ and having a small negative values for $m_{Hu}^2$ at the weak scale and large mixing between top squarks.
The mass spectrum is calculated for the following parameters;
$m_0=5\,$TeV, $m_{1/2}=0.7\,$TeV, $A_0=-8.3\,$TeV, $\tan\beta =10$ with $\mu =0.11\,\text{TeV}$ and $m_A=1\,$\text{TeV}. 
At this benchmark point the neutralino masses are $m_{\tilde{\chi}_{1,2}^0}\approx (101, 118) \;\text{GeV}$, besides of the charginos ($m_{\tilde{\chi}_{1,2}^{\pm}}\approx (113, 611) \;\text{GeV}$) all other particle masses are beyond TeV.

The energy dependence of neutralino pairs of $\tilde{\chi}^0_1\tilde{\chi}^0_1$,  $\tilde{\chi}^0_1\tilde{\chi}^0_2$ and $\tilde{\chi}^0_2\tilde{\chi}^0_2$ in unpolarized photon collisions are given in Fig-\ref{fig:fig1}.
The angular distribution of $\tilde{\chi}^0_1\tilde{\chi}^0_1$,  $\tilde{\chi}^0_1\tilde{\chi}^0_2$ and  $\tilde{\chi}^0_2\tilde{\chi}^0_2$ pairs for two distinct cms energies $\sqrt{s}=0.5\,\text{TeV}-1.0\,\text{TeV}$ are given in Fig-\ref{fig:fig2}-\ref{fig:fig3}.
When the cms energy of the incoming photon is near the masses of neutral $h_0$, $H_0$ and $A_0$ bosons, the s-channel resonance effect such as in triangle diagrams where any of the neutral supersymmetric Higgs bosons are intermediated (Fig-\ref{fig:neupair_tri}) enhances the cross section dramatically.
That effect is seen in each benchmark point investigated.
Therefore, that is also the part where main contribution comes to the total integrated photonic cross section calculated by Eq-\ref{eq:total_cross}.
The peaks seen in Fig-\ref{fig:fig1} around $1\,\text{TeV}$ are due to the intermediation of the neutral Higgs bosons.
The mass spectrum of the neutral Higgs bosons are as following; $m_{h_0/H_0/A_0}=(124.8,1006.7,1000.0)\,\text{GeV}$.
We also take into account the decay widths of these neutral Higgs particles, for that we needed to calculate the decay widths at the nlo-level accuracy using \texttt{FeynHiggs}\cite{Heinemeyer:1998yj,Hahn:2009zz}.
Where the input parameters used in FeynHiggs are set from each benchmark point considered.
In Fig-\ref{fig:fig1a}, it can be seen that the cross section goes up to $1.85\,\text{fb}$ around $\sqrt{s}=384\,\text{GeV}$ for $\tilde{\chi}^0_1\tilde{\chi}^0_1$ pair.
In Fig-\ref{fig:fig1b} the cross section for the $\tilde{\chi}^0_1\tilde{\chi}^0_2$ is drawn where the peaks seen around $\sqrt{s}=1\,\text{TeV}$ are due to the neutral Higgs mediation.
The cross section is enhanced dramatically and it goes up to $0.011\,\text{fb}$.
In Fig-\ref{fig:fig1c}, the cross section for the $\tilde{\chi}^0_2\tilde{\chi}^0_2$ is drawn and it reaches up to $1.95\,\text{fb}$ around $\sqrt{s}=395\,\text{GeV}$.
The angular distribution of each neutralino pair for $0.5\,\text{TeV}$ and $1\,\text{TeV}$ cms energies are given in Fig-\ref{fig:fig2}-\ref{fig:fig3}.
Where at low cms energies there is a small asymmetry at the same order for each neutralino pairs, therefore according to Fig-\ref{fig:fig3} the asymmetry gets large for higher cms energy due to the asymmetrical $t-$ and $u-$terms presented in the cross section. 
Comparing the angular distribution of $\tilde{\chi}^0_1\tilde{\chi}^0_2$ with $\tilde{\chi}^0_1\tilde{\chi}^0_1$ and $\tilde{\chi}^0_2\tilde{\chi}^0_2$ the distribution is close to isotropy.

\item \emph{mSUGRA/CMSSM} : 
Even though the LHC8 ruled out quite large portion of the parameter space there is still a place in the parameter space for dark matter.
This benchmark point employs the following parameters at the GUT scale;
$m_0=10\;\text{TeV}$, $m_{1/2}=0.8 \;\text{TeV}$, $A_0 = -5.45 \;\text{TeV}$ and $\tan\beta= 15$. 
Since the Higgs-doublet mixing parameter $\mu$ is around $\sim235\,\text{GeV}$, the neutralino masses (Eq-\ref{eq:neumix}) at this benchmark point are around $m_{\tilde{\chi}_{1,2}^0}\approx (229, 247) \;\text{GeV}$ and  $m_{\tilde{\chi}_{1,2}^\pm}\approx (248, 701) \;\text{GeV}$ for this benchmark point.
All other sparticles are beyond TeV, therefore, seems like this point is also beyond the reach of LHC.
The energy dependence of the integrated neutralino pair production $\sigma(\gamma\gamma\rightarrow\tilde{\chi}^0_i\tilde{\chi}^0_j)$ for $(i,j)=1,2$ in unpolarized photon collider are depicted in Fig-\ref{fig:fig1}.
In Fig-\ref{fig:fig1a} the neutralino pair production cross section for $\tilde{\chi}^0_1\tilde{\chi}^0_1$  reaches up to $1.0\,\text{fb}$ at $\sqrt{s}=766\,\text{GeV}$.
Whereas the cross section for $\tilde{\chi}^0_2\tilde{\chi}^0_2$ pair is slightly higher and it reaches up to  $1.06\,\text{fb}$ at $\sqrt{s}=774\,\text{GeV}$.
The angular distribution of neutralino pairs at $\sqrt{s}=0.5\,\text{TeV}$ are given in Fig-\ref{fig:fig2}, compared to other benchmark points all three neutralino pairs are flat, the cross section is in complete isotropy.
Therefore at $\sqrt{s}=1\,\text{TeV}$ the angular dependence shows large asymmetry for the same neutralino pairs (Fig-\ref{fig:fig3a},\ref{fig:fig3b}) and it is small for the different neutralino pair (Fig-\ref{fig:fig3c}).

\item \emph{Br\"ummer-Buchm\"ulcer benchmark (BB)} : 
This scenario is proposed by Brummer and Buchmuller \cite{Brummer:2012zc} and it is inspired by GUT-scale string compactifications, where the Fermi scale comes up as a focus point.
In this model the Higgs-doublet mixing parameter $\mu$ comes out from gravitational interactions, in a result it is predicted that graviton mass and $\mu$ are of the same order ($\mu\simeq m_{3/2}\simeq 150-200\,$GeV).
In this study we took the benchmark model specifically adopted for ILC studies, 
where the messenger indices are $(N_1,\ N_2,\ N_3)=(46,46,20)$, $m_{\mathrm{GM}}=250\,$GeV, $\tan\beta = 48$ and Higgs-doublet mixing parameter $\mu=167\,$GeV and $m_A=4.05\,$TeV.
Consequently, the neutralinos $(\tilde{\chi}^0_{1,2})$ and light charginos $(\tilde{\chi}^{\pm}_{1,2})$ are accessible in $\gamma\gamma$ collisions at ILC.
The neutralino and light chargino masses are around $m_{\tilde{\chi}_{1,2}^0}\approx (167, 168) \;\text{GeV}$ and  $m_{\tilde{\chi}_{1}^{\pm}}\approx (167) \;\text{GeV}$, respectively.
The energy dependence of the integrated cross section of $\sigma(\gamma\gamma\rightarrow\tilde{\chi}^0_i\tilde{\chi}^0_j)$ for $(i,j)=1,2$ are depicted in Fig-\ref{fig:fig1}.
The mass splitting between neutralinos is so small, accordingly the cross section for the same neutralino pairs reaches up to $1.43\,\text{fb}$ at $\sqrt{s}=477-538\,\text{GeV}$ for $\tilde{\chi}^0_1\tilde{\chi}^0_1$ and $\tilde{\chi}^0_2\tilde{\chi}^0_2$ pairs.
Therefore, the integrated cross section for the $\tilde{\chi}^0_1\tilde{\chi}^0_2$ is so small, it is negligible.
The angular distribution of neutralino pairs for $\sqrt{s}=0.5\,\text{TeV}-1.0\,\text{TeV}$ are given in Fig-\ref{fig:fig2}-\ref{fig:fig3}, respectively.
The angular distribution for each neutralino pairs at $\sqrt{s}=0.5\,\text{TeV}$ shows a small asymmetry, therefore, at higher cms energy the same neutralino pairs drawn in Fig-\ref{fig:fig3} shows high asymmetry.

\item \emph{Natural Susy (NS)} : 
In this scenario the benchmark point is defined with parameters $m_0(1,2)= 13.35\,\text{TeV}$, $m_0(3) = 0.76\,\text{TeV}$, $m_{1/2} = 1.38\,\text{TeV}$, $A_0 = -0.167 \,\text{TeV}$, $\tan\beta=23$,  $\mu=0.225 \,\text{TeV}$, and $m_A = 1.55 \,\text{TeV}$. 
The neutralino and light chargino masses for this benchmark points are $m_{\tilde{\chi}_{1,2}^0}\approx (224, 232) \;\text{GeV}$ and  $m_{\tilde{\chi}_{1}^{\pm}}\approx (233) \;\text{GeV}$, respectively.
The energy dependence of the integrated $\sigma(\gamma\gamma\rightarrow\tilde{\chi}^0_i\tilde{\chi}^0_j)$ for $(i,j)=1,2$ are depicted in Fig-\ref{fig:fig1}.
The cross section for the same neutralino pairs are close to each other and reaches up to $1.01-1.06\,\text{fb}$, respectively at $\sqrt{s}=766-775\,\text{GeV}$  for $\tilde{\chi}^0_1\tilde{\chi}^0_1$ and $\tilde{\chi}^0_2\tilde{\chi}^0_2$ pairs.
Angular distribution depicted in Fig-\ref{fig:fig2} at $\sqrt{s}=0.5\,\text{TeV}$ shows small variation, meaning almost isotropy in the distribution.
For higher cms energy at $\sqrt{s}=1\,\text{TeV}$, the asymmetry gets higher for the same neutralino pairs.
\end{itemize}

Apart from these benchmark points, the calculation is also carried out for the NUHM2\footnote{In the two-parameter non-universal Higgs model (NUHM2), the soft-breaking scalar masses of the two Higgs doublets $\hat{H}_u$ and $\hat{H}_d$ are free parameters, they are independent of the sfermion masses.
Since the $\hat{H}_u$ and $\hat{H}_d$ fields are not belong to the same multiplets, there is no driving force to assume the Higgs fields and sfermion fields unifiy.
In this model all Higgs bosons are light whereas the rest of the sparticles are beyond current LHC reach.} 
and NUGM\footnote{
This benchmark point, the non-universal gaugino masses (NUGM), inspired by the GUT models, where 
the universality of the gauging masses are relaxed at $M_{GUT}$ as a result it helps to resolve the little hierarchy problem.} presented in \cite{Baer:2013ula}. 
However, considering the expected total luminosity of the $\gamma\gamma$ collisions, cross sections for the lightest neutralino pair which are less than attobar are not presented.
Due to the very small production rates, the lightest neutralino pairs are not accessible in $\gamma\gamma$ collider for these benchmark points.
Contrary to that, the second lightest neutralino pair production is substantially high at the order of $10\,\text{fb}$ in these two benchmark points.
The total integrated photonic cross section values of each neutralino pairs for the benchmark points given above are presented in Table-\ref{tab:photcross}.
The cross sections where it is not kinematically possible at a given cms for the $\gamma\gamma$ collisions in ILC are not included in the table.

\begin{table}[htdp]
\caption{
Integrated total photonic cross section of $e^+e^-\rightarrow\gamma\gamma\rightarrow\tilde\chi_i^0\tilde\chi_j^0$ 
for various benchmark points at cms energy $\sqrt{s}=0.5-1.0\;TeV$.}
\begin{center}
\begin{ruledtabular}
\begin{tabular}{c|ccc}
Benchmark			&	$\tilde{\chi}_1^0\tilde{\chi}_1^0\;\;\;[fb]$	&			$\tilde{\chi}_1^0 \tilde{\chi}_2^0\;\;\;[fb]$	&		$\tilde{\chi}_2^0\tilde{\chi}_2^0\;\;\;[fb]$	\\
Scenario		&	$\sqrt{s}=0.5/1\,\text{TeV}$			&			$\sqrt{s}=0.5/1\,\text{TeV}$				&		$\sqrt{s}=0.5/1\,\text{TeV}$		\\
\hline
RNS			&	0.747/1.258 	&	0.002/0.005	&	0.634/1.229	\\
NS			&	 	-/0.577 	&	 -/0.005		&	 -/0.410 		\\
mSUGRA		&	 	-/0.340 	&	-/$<10^{-3}$	&	-/0.339		\\
BB			&	 0.124/0.809 	&	$<10^{-5}$	&	0.120/0.808	\\
\end{tabular}
\end{ruledtabular}
\end{center}
\label{tab:photcross}
\end{table}%

\section{Conclusion}
\label{sec:5}

In this paper we have calculated the production rates of different neutralino pairs at the NLO level including all possible Feynman diagrams in a future photon collider.
The numerical analysis on production rates as a function of cms energy, angular dependence and the total integrated photonic cross section are calculated at the ILC with four different susy benchmark points.
These benchmark points had been introduced in the face of new susy constraints set by resent LHC data.
In addition to that, these points are reachable, particularly, at the International Linear Collider.
This study evaluates the neutralino pair production rates for the benchmark points presented.
The production rates of $\tilde{\chi}_1\tilde{\chi}_1$,  $\tilde{\chi}_1\tilde{\chi}_2$ and $\tilde{\chi}_2\tilde{\chi}_2$ pairs via photon collisions are 
analyzed up to $\sqrt{s}=1.1\,\text{TeV}$.
The peaks seen in the distributions are due to resonance effect of neutral Higgs mediation in triangle diagrams.
Among the four benchmark points, the RNS gives the highest production rates for the same neutralino pairs and highest forward-backward peaks in the distribution.
However, the production rates are still comparable among the benchmark points.
Meanwhile, the BB benchmark point produces the smallest rates for different neutralino pair. 

The asymmetry in the production of the each neutralino pairs among the benchmark points are analyzed by the angular distribution for two distinct cms energies at $\sqrt{s}=0.5\,\text{TeV}$ and $\sqrt{s}=1\,\text{TeV}$.
MS and NS benchmark points show near full isotropy in forward-backward scattering at $\sqrt{s}=0.5\,\text{TeV}$ for each neutralino pairs.
However, RNS and BB benchmark points develops moderate asymmetry at the same cms energy for each neuralino pairs.
Moving at higher cms energy, the isotropy breaks and production rates for the same neutralino pairs develops quite large forward-backward scattering in a result the asymmetry is detectable by large fraction.
Compared to the lower cms energy, the asymmetry presents itself due to asymmetrical $u$- and $t$-terms in the amplitudes.

The integrated total photonics cross section is calculated by convoluting the $\gamma\gamma \rightarrow \tilde{\chi}_i\tilde{\chi}_j$ cross section with the photon luminosities at the ILC for two distinct cms energies of the incoming $e^+e^-$ beams.
Since the neutralinos, according to R-parity conservation are the lowest massive supersymmetric particles which could be produced in a decay of a supersymmetric particle, are natural candidate for the dark matter.
If LHC in its lifetime couldn't find any hint on BSM specifically supersymmetry, benchmark scenarios or nice portion of the parameter space would be ruled out.
Therefore, some space on supersymmetry parameters would be left out and a future Linear Collider in $\gamma\gamma$ collision mode 
could rule out these left spaces or discover the BSM signals.
ILC in $\gamma\gamma$ collision mode have very small experimental background, since the neutralinos would produce a large missing energy in the collisions, production of missing energy with the asymmetry distribution in the events and nice acceptance for the detectors would give the hints of supersymmetry.
The results concludes that a photon collider with an additional very small cost compared to the $e^+e^-$ collider would produce new results and that would show the secrets of our universe.


\section{Acknowledgement}
It is a pleasure to thank T.Hahn for fruitful discussions and his time.

\end{document}